\documentclass[preprint,12pt]{elsarticle}
\usepackage{graphicx}
\usepackage{graphicx}
\usepackage{lipsum}
\usepackage{amssymb}
\usepackage{amsmath}
\usepackage{color}
\usepackage{multirow}
\usepackage{afterpage}
\usepackage{dsfont}
\usepackage{bm}%
\def\fnum@figure{\figurename\thefigure}
\renewcommand{\figurename}{Fig.}

\journal{Physica A}
\begin{document}

\title
 {Fidelity based Measurement Induced Nonlocality and its dynamics in Quantum noisy channels}
\author{R. Muthuganesan, R. Sankaranarayanan}
\address{Department of Physics, National Institute of Technology\\ Tiruchirappalli-620015, Tamil Nadu, India.}

\begin{abstract}
Measurement induced nonlocality (MIN) captures global nonlocal effect of  bipartite quantum state due to locally invariant projective measurements. In this paper, we propose a new version of MIN using fidelity induced metric,and the same is calculated for pure and mixed states. For mixed state, the upper bound is obtained from eigenvalues of correlation matrix. Further, dynamics of MIN and fidelity based MIN under various noisy quantum channels show that they are more robust than entanglement.


\end{abstract}

\maketitle
\section{Introduction}
One of the most intriguing features of quantum regime is that local measurement on a part of composite system can induce global influence on the system. Such influence, also called as nonlocality, has no analogue in the classical scale. The strange non-classical phenomenon is attributed to correlation between different parts of the system. Understanding the correlation of simplest composite system, namely bipartite system, is fundamental and relevant for quantum information theory. In this context, many measures of correlation for bipartite system have been proposed in recent years. One notable measure, which goes beyond entanglement, is quantum discord as proposed by Ollivier and Zurek \cite{Ollivier2001}. Though the computation of discord involves complex optimization procedure \cite{Girolami2011}, non-zero discord of separable states reveals that entanglement is not a complete manifestation of nonlocality or quantum correlation.

It is well known for pure states that while separable sates are invariant under von Neumann projective (local) measurements, inseparable states are altered by such measurements. Hence, local measurements may be useful tool for quantifying quantum correlation. On the other hand, notion of geometric quantum discord - distance between an arbitrary state and closest zero discord state, was introduced as a measure of correlation \cite{Dakic2010}. This notion is conveniently reformulated as the minimized square of Hilbert-Schmidt norm of difference between pre- and post- projective measurement of state under consideration \cite{Luo2010pra}. Further, Luo and Fu presented a new measure of nonlocality for bipartite system, which is also dual to geometric discord, termed as measurement induced nonlocality (MIN) \cite{Luo2011}. Both the quantities are significant figure of merit for quantum correlations with wide applications \cite{Wiseman2007,Peters2005, Mattle1996}.

However, both the quantities suffer from the so called local ancilla problem - change may be effected through some trivial and uncorrelated action of the unmeasured party \cite{Piani2012}. This problem can be circumvented by replacing density matrix by its square root \cite{Chang2013}. Based on this, MIN has also been investigated in terms of relative entropy \cite{Xi2012}, von Neumann entropy \cite{Hu2012}, skew information \cite{Li2016} and trace distance \cite{Hu2015}. Further, MIN has been investigated for bound entangled states \cite{Rana2013}, general bipartite system \cite{Mirafzali2011} and Heisenberg spin chains \cite{Chen2015,Muthuganesan2017}. The dynamics and monogamy of measurement induced nonlocality has also been studied \cite{Sen2012,Sen2013}.

In this article, we introduce fidelity based measurement induced nonlocality to extract nonlocal effects of two qubit states due to projective measurements. It is shown that this quantity is naturally remedying the local ancilla problem of  MIN and also easy to measure. Since fidelity is also experimentally accessible using quantum networks \cite{Miszczak2009}, nonlocal measure based on fidelity also enjoys physical relevance. For pure state, we show that the fidelity based MIN is indeed coinciding with other forms of MIN (Hilbert-Schmidt norm, skew information), and geometric discord. Our investigations also provide a closed formula for $2\times n$ dimensional mixed state and an upper bound for arbitrary $m\times n$ dimensional mixed state. Further, we study the dynamics of MIN  and fidelity based MIN under various noisy channel such as amplitude damping, depolarizing and generalized amplitude damping. It is shown that the MINs are robust measures of quantum correlation than entanglement against decoherence induced by the noisy channels.

\section{MIN based on Fidelity}
Let us consider a bipartite quantum state $\rho $ in a Hilbert space $\mathcal{H}^a\otimes \mathcal{H}^b$. MIN is defined as the square of Hilbert-Schmidt norm of difference between pre- and post-measurement state i.e., \cite{Luo2011}
\begin{equation}
 N(\rho ) =~^{max}_{\Pi ^{a}}\| \rho - \Pi ^{a}(\rho )\| ^{2} 
\end{equation}
where the maximum is taken over the von Neumann projective measurements on subsystem $a$. Here $\Pi^{a}(\rho) = \sum _{k} (\Pi ^{a}_{k} \otimes   \mathds{1} ^{b}) \rho (\Pi ^{a}_{k} \otimes    \mathds{1}^{b} )$, with $\Pi ^{a}= \{\Pi ^{a}_{k}\}= \{|k\rangle \langle k|\}$ being the projective measurements on the subsystem $a$, which do not change the marginal state $\rho^{a}$ locally i.e., $\Pi ^{a}(\rho^{a})=\rho ^{a}$. In fact, the MIN has a closed formula for $2\times n$ dimensional states.

However, Hilbert-Schmidt norm based MIN could change due to  trivial and uncorrelated action on the unmeasured party $b$. This arises from appending an uncorrelated ancilla $c$ and regarding the state $\rho ^{a:bc}=\rho^{ab} \otimes  \rho ^{c}$ as a bipartite state with the partition $a$:$bc$; then it is easy to verify the following 
\begin{equation}
 N(\rho^{a:bc} ) = N(\rho^{ab})tr(\rho ^{c})^2.  \nonumber  
\end{equation}
This relation implies that as long as $\rho^{c}$ is a mixed, MIN is altered by the addition of uncorrelated ancilla $c$ - local ancilla problem.

  We can resolve local ancilla problem by defining MIN based on fidelity, which is a measure of closeness between two arbitrary states $\rho $ and $\sigma $. Defining fidelity as $F(\rho,\sigma )= \left(tr\sqrt{\sqrt{\rho }\sigma \sqrt{\rho }} \right)^{2}$, one can define a metric $D(\rho,\sigma )=\Phi (F(\rho,\sigma))$, where $\Phi$ is a monotonically decreasing function of $F(\rho,\sigma )$ and $\Phi$ is required to satisfy all the axioms of distance measure \cite{Jozsa1994}. Due to computational complexity of fidelity, in what follows we employ an alternate definition of fidelity as \cite{Wang2008}
\begin{equation}
\mathcal {F}(\rho ,\sigma )=\frac{(tr(\rho \sigma ))^2}{tr(\rho^{2})~tr(\sigma^{2} )} \nonumber   
\end{equation}
which satisfies all the axioms to measure the closeness of two states. Defining MIN in terms of fidelity induced metric (F-MIN) as
\begin{equation}
N_{\mathcal {F}}(\rho ) =~^{max}_{\Pi ^{a}}\mathcal {C}^{2}(\rho, \Pi ^{a}(\rho )) \label{min}  
\end{equation}
where  $\mathcal {C}(\rho ,\sigma )=\sqrt{1-\mathcal {F}(\rho ,\sigma )}$ is sine metric. In other words, MIN is defined in terms of the fidelity between pre- and post- measurement state. This quantity can remedy the local ancilla problem of MIN as shown below. After the addition of local ancilla the fidelity between the pre- and post- measurement state is 
\begin{equation}
 \mathcal {F}\left(\rho^{a:bc},\Pi ^{a}(\rho^{a:bc})\right) = \mathcal {F}\left(\rho^{ab}\otimes  \rho ^{c},\Pi ^{a}(\rho^{ab})\otimes  \rho ^{c}\right). \nonumber
\end{equation}
Using multiplicativity property of fidelity,
\begin{equation}
 \mathcal {F}\left(\rho^{a:bc},\Pi ^{a}(\rho^{a:bc})\right) = \mathcal {F}\left(\rho^{ab}, \Pi ^{a} (\rho ^{ab})\right).~\mathcal {F}(\rho^{c},\rho^{c})=\mathcal {F}\left(\rho^{ab},\Pi ^{a}(\rho^{ab})\right)  \nonumber
\end{equation}
Hence $N_{\mathcal {F}}(\rho)$ is a good measure of nonlocality or quantumness in a given system.
\section{MIN for pure state}
\label{sec:2}
{\bf Theorem 1:} For pure bipartite state with Schmidt decomposition $| \Psi \rangle =\sum_{i}\sqrt{\lambda _{i}}| \alpha _{i} \rangle \otimes | \beta _{i}\rangle $  F-MIN is 
\begin{equation}
N_{\mathcal {F}}(| \Psi \rangle\langle \Psi| )=1- \sum_{i}\lambda_{i}^{2}. \label{eq:pure}
\end{equation}
The proof is as follows. The von Neumann projective measurement on party $a$ is expressed as $\Pi ^{a}=\{\Pi ^{a}_{k}\} = \{| \alpha_{k}\rangle \langle \alpha_{k}|\} $. The projective measurements do not alter the marginal states i.e., $\left(\Pi ^{a}(\rho^{a})=\sum_{k}\Pi ^{a}_{k}\rho^{a}\Pi ^{a}_{k}=\rho^{a}\right)$. Noting that
\begin{equation}
\rho=| \Psi \rangle \langle \Psi| = \sum_{ij}\sqrt{\lambda_{i}\lambda_{j}}| \alpha_{i} \rangle \langle \alpha_{j}| \otimes  | \beta_{i} \rangle \langle \beta_{j}|. \nonumber  
\end{equation}
Since $tr(\rho~ \Pi ^{a}(\rho))=tr(\Pi ^{a}(\rho))^2$ the fidelity between pre- and post- measurement state becomes
\begin{equation}
\mathcal {F}(\rho ,\Pi ^{a}(\rho))=\sum_{k}\lambda ^{2}_{k} \nonumber  
\end{equation}
and hence the theorem is proved. Thus F-MIN coincides with Hilbert-Schmidt norm \cite{Luo2011} and skew information \cite{Li2016} based MINs and geometric discord \cite{Luo2010pra} for pure states.
\section{MIN for mixed state}
Let $\{X_{i}:i=0,1,2,\cdots,m^{2}-1\} \in \mathcal{B}(\mathcal{H}^a)$ be a set of orthonormal operators for the state space $\mathcal{H}^a$ with operator inner product $\langle X_{i}| X_{j}\rangle = tr(X_{i}^{\dagger}X_{j})$. Similarly, one can define $\{Y_{j}:j=0,1,2,\cdots,n^{2}-1\}  \in \mathcal{B}(\mathcal{H}^b)$ for the state space of $\mathcal{H}^b$. The operators $X_{i}$ and $Y_{j}$ are satisfying the conditions $tr(X_{k}^{\dagger }X_{l})=tr(Y_{k}^{\dagger}Y_{l})=\delta _{kl}$. With this one can construct a set of orthonormal operators $\{X_{i} \otimes Y_{j} \}\in \mathcal{B} (\mathcal{H}^{a}\otimes \mathcal{H}^{b}) $ for the composite system. Consequently, an arbitrary state of a bipartite composite system can be written as
\begin{equation}
\rho= \sum_{i,j}\gamma _{ij}X_{i}\otimes  Y_{j} \label{c}
\end{equation}
where $\Gamma = (\gamma _{ij} =tr (\rho ~X_{i}\otimes  Y_{j}))$ is a $m^2 \times n^2$ real matrix.

After a straight forward calculation, the fidelity between pre- and post- measurement state is computed as
\begin{equation}
\mathcal{F}(\rho,\Pi^{a}(\rho) )=\frac{tr(A\Gamma \Gamma ^{t}A^{t})}{\| \Gamma \|^{2} } \nonumber
\end{equation}
where the matrix $A=(a_{ki}=tr(|k \rangle \langle k| X_{i}))$ is a rectangular matrix of order $m\times m^{2}$. Then, F-MIN is 
\begin{equation}
N_{\mathcal{F}}(\rho)=\frac{1}{\|\Gamma  \|^{2} }\left[\|\Gamma  \|^{2}-^{min}_{A}tr(A\Gamma \Gamma ^{t} A^{t})\right]. \label{result}
\end{equation}

{\bf Theorem 2:} F-MIN has a tight upper bound as 
\begin{equation}
N_{\mathcal{F}}(\rho)\leq \frac{1}{\|\Gamma  \|^{2} }\left(\sum_{i=m}^{m^2-1}\mu _{i}\right) \label{FMIN}
\end{equation}
where $\mu_{i}$ are eigenvalues of matrix ${\bf xx}^{t}+TT^{t}$, derived from $\Gamma $,  listed in increasing order.

Adapting the method \cite{SLuo2012}, we prove the theorem as follows: If $X_{0}=\mathds{1}^{a}/\sqrt{m}$, $Y_{0}=\mathds{1}^{b}/\sqrt{n}$, and separating the terms in eq.(\ref{c}), the state $\rho$ can be written as  
\begin{equation}
\rho =\frac{1}{\sqrt{m n}}\frac{\mathds{1}^{a}}{\sqrt{m}}\otimes \frac{\mathds{1}^{b}}{\sqrt{n}}+\sum_{i=1}^{m^2 -1}x_{i}X_{i}\otimes\frac{\mathds{1}^{b}}{\sqrt{n}}+\frac{\mathds{1}^{a}}{\sqrt{m}}\otimes\sum_{j=1}^{n^2 -1}y_{j}Y_{j} +\sum_{i,j\neq 0}t_{ij }X_{i}\otimes Y_{j} \label{rho6}
\end{equation}
where   $x_{i}=tr(\rho ~X_{i}\otimes \mathds{1}^{b})/\sqrt{n}$, $y_{j}=tr(\rho ~\mathds{1}^{a} \otimes Y_{j} )/\sqrt{m}$ and $T = (t _{ij} =tr (\rho ~X_{i}\otimes  Y_{j}))$ is a real correlation matrix of order $(m^{2}-1)\times(n^2 -1)$.
Comparing eqs.(\ref{c}) and (\ref{rho6}) we obtain $\gamma _{00}=\frac{1}{\sqrt{mn}}$, $\gamma _{i0}=x_{i}$, $\gamma _{0j}=y_{j}$, $\gamma _{ij}=t_{ij}$ with $i=1,2,3,\cdots, m^2 -1 $ and $j=1,2,3,\cdots,n^2 -1$. With this we write $\Gamma $ matrix as   
\begin{equation}
\Gamma =
\begin{pmatrix}
\frac{1}{\sqrt{mn}} & {\bf y}^{t} \\
{\bf x} & T
\end{pmatrix}  \nonumber
\end{equation}
where ${\bf x}=(x_{1}~x_{2}~x_{3} ~\cdots ~x_{m^2-1})^t$ and ${\bf y}=(y_{1}~y_{2} ~y_{3}~ \cdots ~y_{n^2-1})^t$. It is easy to show that

\begin{equation}
\| \Gamma \|^{2}=tr(\Gamma \Gamma ^{t})=\frac{1}{mn}+{\bf y}^{t}{\bf y}+tr({\bf x}{\bf x}^t + TT^{t}). \label{eq:8}
\end{equation}

Let us define a vector
\begin{equation}
{\bf a}_{k}=\sqrt{\frac{m}{m-1}}\left(a_{k1}~a_{k2}~a_{k3}~\cdots~ a_{k(m^{2}-1)}\right)^{t} 
\end{equation} 
which satisfies $\| {\bf a}_{k} \|^{2}=1$, ${\bf a}_{k}^{t}{\bf a}_{k'}=-1/(m-1)$, and $\sum _{k=1}^{m}a_{ki}=0$. With this setting an $m\times(m^2 -1)$ matrix
\begin{equation}
A_{0}=
\begin{pmatrix}
{\bf a}_{1}^{t}\\
{\bf a}_{2}^{t}\\
\vdots \\
{\bf a}_{m}^{t}
\end{pmatrix}
\end{equation} 
and using the properties of ${\bf a}_{k}$, we write
\begin{equation}
A_{0}A_{0}^{t}=
\begin{pmatrix}
1 & -1/(m-1) & \cdots & -1/(m-1) \\
-1/(m-1) & 1 &\cdots & -1/(m-1)\\
\vdots & \vdots & \ddots &\vdots \\
-1/(m-1) & -1/(m-1) & \cdots & 1 
\end{pmatrix}. \nonumber
\end{equation} 
Noting that,
\begin{equation}
A==\frac{1}{\sqrt{m}}
\begin{pmatrix}
1 & \sqrt{(m-1)}{\bf a}_{1}^{t}\\
1 & \sqrt{(m-1)}{\bf a}_{2}^{t}\\
\vdots & \vdots \\
1 & \sqrt{(m-1)}{\bf a}_{m}^{t}
\end{pmatrix} \nonumber
\end{equation} 
then the direct multiplication gives
\begin{equation}
^{min}_{A}~  tr(A\Gamma \Gamma ^{t}A^{t})=\frac{1}{mn}+{\bf y}^t{\bf y}+\frac{m-1}{m}~^{min}_{A_{0}}~tr(A_{0}({\bf x}{\bf x}^t + TT^{t})A_{0}^{t}). \label{eq:11}
\end{equation}
Here $A_{0} A_{0}^{t}$ is a real symmetric matrix with eigenvalues $0$ and $m/(m-1) ~(m-1~ \text{times})$. Let us consider a similarity transformation $A_{0} A_{0}^{t}=UDU^{t}$, where $U$ is an orthogonal matrix and $D$ is a diagonal matrix. Defining $D$ as     
\begin{equation}
D=
\begin{pmatrix}
D_{0} & 0\\
0 & 0
\end{pmatrix} \nonumber
\end{equation} 
where $D_{0}$ is a diagonal matrix of order $m-1$ with entries $m/(m-1)$. Now constructing $m \times m^{2}$ matrix $B$ as,
\begin{equation}
B=
\begin{pmatrix}
D_{0}^{-1/2} & 0\\
0 & 0
\end{pmatrix}
U^{t}A_{0}=
\begin{pmatrix}
R\\
0
\end{pmatrix}
 \nonumber
\end{equation}  
where $R$ is $(m-1)\times m^{2}$ matrix, such that $RR^{t}=\mathds{1}_{m-1}$. From the definition of $B$, we have
\begin{equation}
A_{0}=U
\begin{pmatrix}
D_{0}^{-1/2} & 0\\
0 & 0
\end{pmatrix} B. \nonumber
\end{equation} 

After straight forward multiplication and simplification, we obtain
\begin{equation}
^{min}_{A_{0}}~tr(A_{0}({\bf x}{\bf x}^t + TT^{t})A_{0}^{t})=\sum _{i=1}^{m-1}\mu _{i} \label{eq:12}
\end{equation}
where $\mu _{i}$ are eigenvalues of matrix ${\bf x}{\bf x}^t + TT^{t}$ listed in increasing order. Then from eqs.(\ref{eq:8}), (\ref{eq:11}) and (\ref{eq:12}) F-MIN is 
\begin{equation}
N_{\mathcal{F}}(\rho)=\frac{1}{\|\Gamma  \|^{2} }\left[\sum_{i=1}^{m^2-1}\mu _{i} -\sum_{i=1}^{m-1}\mu _{i}\right] \nonumber
\end{equation}
which leads to the upper bound for F-MIN as
\begin{equation}
N_{\mathcal{F}}(\rho)\leq \frac{1}{\|\Gamma  \|^{2} }\left(\sum_{i=m}^{m^2-1}\mu _{i}\right) \nonumber
\end{equation}
to complete the proof. For a special case of $2 \times n$ dimensional system, we have the optimization as,
\begin{equation}
^{min}_{A}~tr(A\Gamma \Gamma^{t}A^{t})=\mu _{1} \nonumber
\end{equation}
and F-MIN is
\begin{equation}
N_{\mathcal{F}}(\rho)= \frac{1}{\|\Gamma  \|^{2} }\left(\mu _{2}+\mu _{3}\right). 
\end{equation}
\section{Examples}
Here we study the F-MIN and MIN \cite{Luo2010} for two well-known families of mixed state namely, isotropic state and Werner state.

1. First we consider $m\times m$ dimensional isotropic state in the form \cite{Horodecki1999}
\begin{align}
\rho ^{ab}=\frac{1-x}{m^{2}-1}\mathds{1}+\frac{m^{2}x -1}{m^{2}-1}| \Psi^{+} \rangle \langle \Psi^{+} |
\end{align} 
where $| \Psi^{+}\rangle=\frac{1}{\sqrt{m}}\sum_{i}|ii \rangle$, $\mathds{1}$ is identity matrix of order $m^{2}\times m^{2}$ and $x\in [0,1]$. From eq.(\ref{min}) the fidelity based MIN for this state is computed as
\begin{align}
N_{\mathcal{F}}(\rho ^{ab})=\frac{(m^{2} x-1)^{2}}{m^2(1-x)^{2}+(m-1)(1+m x)^{2}+(m^{2}x-1)^{2}}.
\end{align}
This result shows that the F-MIN vanishes only when  $x=1/m^{2}$, at which $\rho^{ab}=\mathds{1}/m^{2}$ being the maximally mixed state. Further, in the asymptotic limit
\begin{equation}
\lim_{m \rightarrow \infty } N_{\mathcal{F}}(\rho ^{ab})=1~\text{and} \lim_{m \rightarrow \infty } N(\rho ^{ab})=x^2. \nonumber
\end{equation}

2. Next we consider $m\times m$ dimensional Werner state \cite{Werner1989}   
\begin{equation}
\omega ^{ab}=\frac{m-x}{m^{3}-m}\mathds{1}+\frac{m x -1}{m^{3}-m}P
\end{equation}
where $P=\sum_{\alpha ,\beta }| \alpha \rangle \langle \beta | \otimes  |\beta\rangle \langle \alpha|$ is flip operator with $x\in [-1,1]$. From eq.(\ref{min}) we compute F-MIN as,
\begin{equation}
N_{\mathcal{F}}(\omega  ^{ab})=\frac{(m x-1)^{2}}{(m-x)^{2}+(m-1)(x+1)^{2}+(m x-1)^{2}}.
\end{equation}
This result shows that  F-MIN vanishes only when $x=1/m$, at which $\omega  ^{ab}=\mathds{1}/m^{2}$ being the maximally mixed state. In the asymptotic limit we have,
\begin{equation}
\lim_{m \rightarrow \infty } N_{\mathcal{F}}(\omega  ^{ab})=\frac{x^2}{1+x^2}~\text{and} \lim_{m \rightarrow \infty } N(\omega ^{ab})=0. \nonumber
\end{equation}
\section{Dynamics of MIN}
In this section, we study the dynamics of MIN in various dissipative quantum channels such as amplitude damping, depolarizing and generalized amplitude damping. Considering an initial state for two qubits $\rho (0)$, its evolution can be written as \cite{Nielsen2010}
\begin{equation}
\rho(t)=\sum_{i,j}E_{i,j}\rho (0)E_{i,j}^{\dagger }
\end{equation}
with Kraus operators $E_{i,j}=E_{i}\otimes E_{j}$ satisfying the condition $\sum_{i,j}E_{i,j}^{\dagger } E_{i,j}=\mathds{1}$. In what follows we consider a family of pure entangled states $| \psi \rangle = \sqrt{\alpha }|00 \rangle +\sqrt{1-\alpha }|11 \rangle $ with $\alpha \in [0,1]$ as the initial state, i.e., $\rho (0)=| \psi\rangle \langle  \psi | $. For $\alpha =1/2$, $| \psi\rangle $ is maximally entangled state.

\textit{(i) Amplitude damping:} 
First of all we consider the time evolution of initial state via amplitude damping channel, which is a classical noise process describing dissipative interaction between the system and environment. There is an exchange of energy between system and environment, such that system is driven into thermal equilibrium with environment. This channel may be modelled by treating environment as a large collection of independent harmonic oscillators interacting weakly with system, as in the case of the spontaneous emission of an excited atom in the vacuum electromagnetic field (the reservoirs are at zero temperature, i.e., in the vacuum state). The Kraus operators for a single qubit amplitude damping are given by \cite{Nielsen2010}
\begin{equation}
E_{0}=
\begin{pmatrix}
1 & 0\\
0 & \sqrt{1-\gamma }
\end{pmatrix}
\text{and} ~E_{1}=
\begin{pmatrix}
0 & \sqrt{\gamma }\\
0 & 0
\end{pmatrix}
\end{equation} 
where $\gamma =1-\mathrm{e}^{-\gamma' t}$ and $\gamma'$ is the decay rate or rate of spontaneous emission. The parameter $\gamma $ plays the role of time such that $0\leq t\leq \infty $ mapped on to $0\leq \gamma \leq 1$. The matrix elements of evolved state are 
\begin{eqnarray}
\rho _{11}(t)&=&\rho _{11}(0),\nonumber \\ 
\rho _{22}(t)&=&\rho _{33}(t)=\rho _{14}(0)\gamma (1-\gamma ),\nonumber \\
\rho _{44}(t)&=&\rho _{44}(0)(1-\gamma )^{2},\nonumber \\
\rho _{14}(t)&=&\rho _{41}(t)=\rho _{14}(0)(1-\gamma ). \nonumber 
\end{eqnarray}
with remaining elements of density matrix being zero. The concurrence \cite{Hill1997} as measure of entanglement, Hilbert-Schmidt norm and fidelity based MINs for evolved state are given as
\begin{figure*}[!ht]
\centering\includegraphics[width=0.8\linewidth]{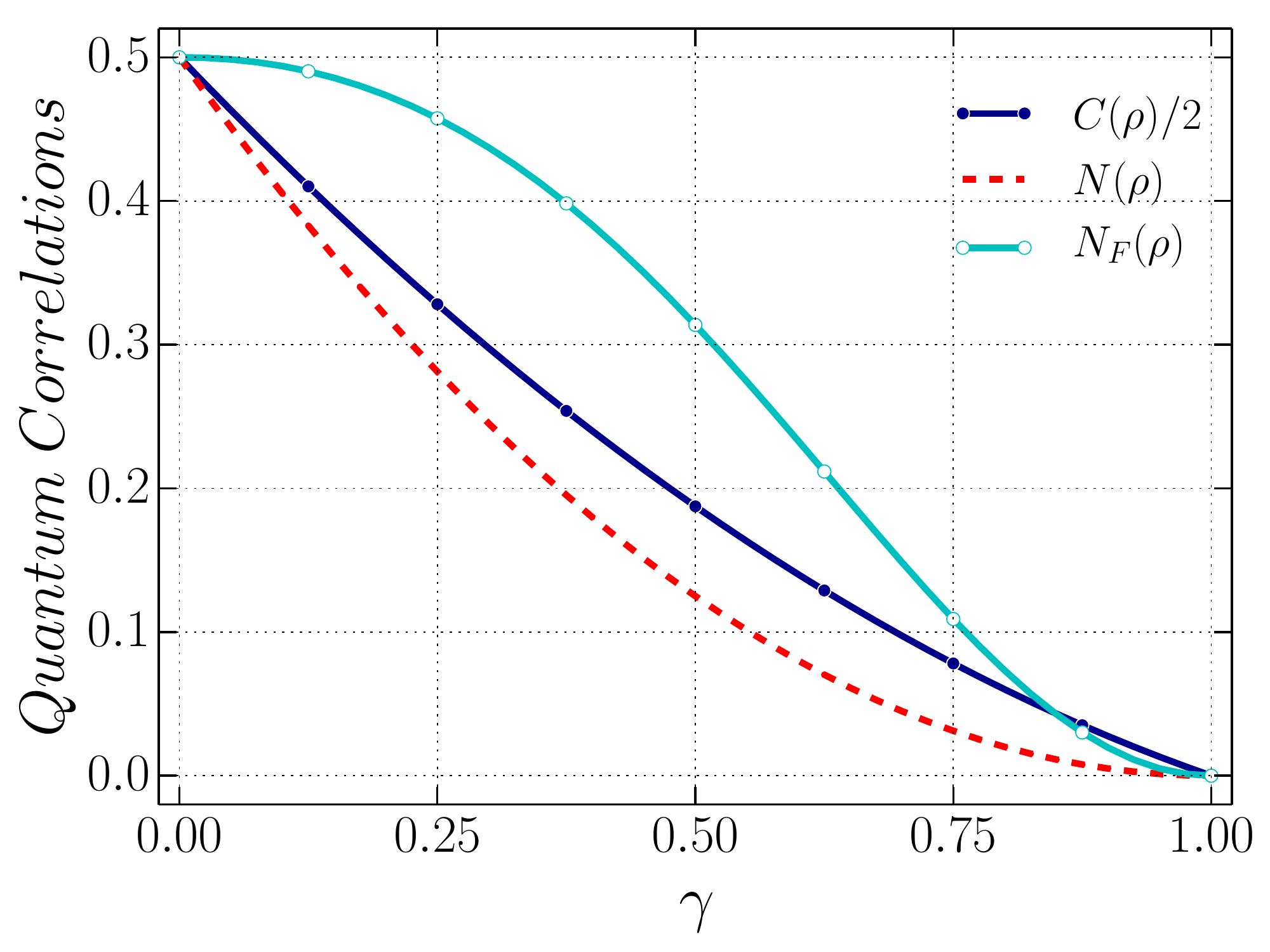}
\caption{(color online) Dynamics of concurrence, Hilbert- Schmidt norm and fidelity based measurement induced nonlocalities under amplitude damping channel for the initial state $| \psi \rangle =\frac{1}{\sqrt{2}}(| 00 \rangle + |11 \rangle ) $.}
\label{fig1}
\end{figure*}
\begin{eqnarray}
 \label{eq:HS} 
C(\rho (t))/2 &=& max~\{0,\rho _{14}(t)-\rho _{22}(t)\}, \nonumber \\
N(\rho (t)) &=& 2\rho_{14}(t)^{2}, \\
N_{\mathcal{F}}(\rho (t)) &=&\frac{2\rho _{14}(t)^{2}}{\rho _{11}(t)^{2}+2\rho _{14}(t)^{2}+2\rho _{22}(t)^{2}+\rho _{44}(t)^{2}}.  \nonumber 
\end{eqnarray}
Since $\rho _{14}(0)=\sqrt{\alpha (1-\alpha )}$, from the above results we understand that concurrence and both the forms of MIN vanish identically for the initial product states $\alpha = 0,1$. Further, it is also clear that for all other initial entangled states the concurrence, MIN and F-MIN vanish for $\gamma =1$. In other words, the entanglement and nonlocal (quantum) correlation disappear asymptotically due to the amplitude damping channel. As an example, we have plotted the dynamics of MINs and concurrence for the initial state $\alpha = 1/2$ in Fig.\ref{fig1}. It is observed that concurrence, MIN and F-MIN decrease with the increase of $\gamma $ and they vanish at $\gamma = 1$.

\begin{figure*}[!ht]
\centering\includegraphics[width=0.8\linewidth]{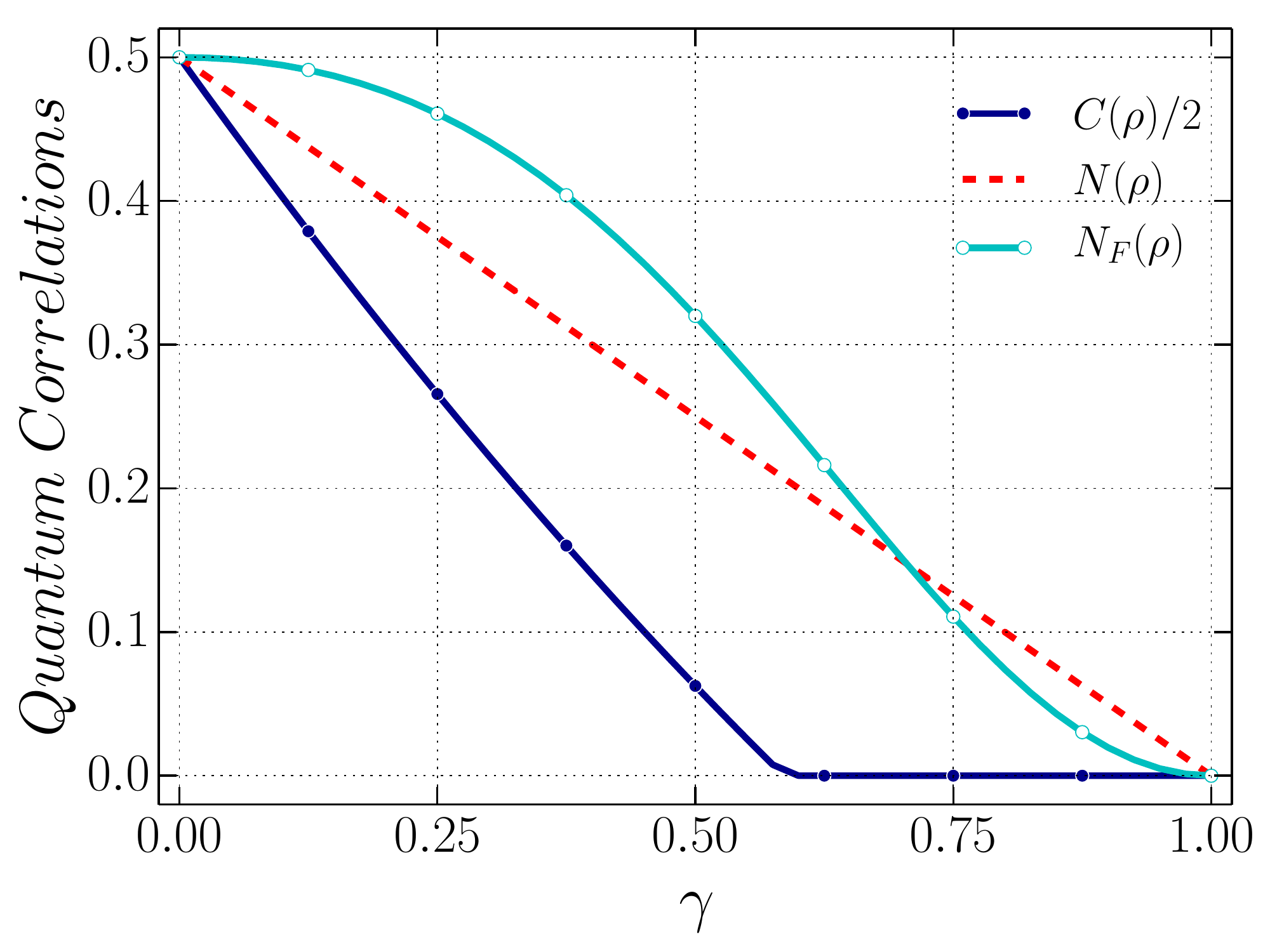}
\caption{(color online) Dynamics of concurrence, Hilbert-Schmidt norm and fidelity based measurement induced nonlocalities under depolarizing channel for the initial state $| \psi \rangle =\frac{1}{\sqrt{2}}(| 00 \rangle + |11 \rangle )$.}
\label{fig2}
\end{figure*}

\textit{(ii) Depolarizing:}
This channel is a type of quantum noise which transforms a single qubit into a maximally mixed state $\mathds{1}/2$ with probability $\gamma $. This channel is represented by the Kraus operators \cite{Nielsen2010}: 
\begin{eqnarray}
E_{0}&=&\sqrt{1-3\gamma /4} ~\mathds{1}, ~E_{1}=\sqrt{\gamma} ~\sigma _{x}/2  \nonumber\\ 
E_{2}&=&\sqrt{\gamma}~ \sigma _{y}/2,~E_{3}=\sqrt{\gamma}~ \sigma _{z}/2  \nonumber
\end{eqnarray}
where $\sigma _{i}$ are Pauli spin matrices and $\gamma =1-\mathrm{e}^{-\gamma't}$ with $\gamma'$ being damping constant. For the above mentioned initial state, density matrix of evolved state assumes the same form as that of amplitude damping with following non-zero matrix elements:
\begin{eqnarray}
\rho _{11}(t)&=&\rho _{11}(0)(1-\gamma )+\gamma ^{2}/4,\nonumber \\
\rho _{22}(t)&=&\rho _{33}(t)=\gamma (1-\gamma /2)/2,\nonumber \\
\rho _{44}(t)&=&1-2\rho _{22}(t)-\rho _{11}(t),\nonumber \\
\rho _{14}(t)&=&\rho _{41}(t)=\rho _{14}(0)(1-\gamma ).\nonumber
\end{eqnarray}
Up on substituting the density matrix elements in eq. (\ref{eq:HS}) we obtain concurrence, MIN and F-MIN for this channel. Here also we find that both the forms of MIN vanish identically for the initial product states $\alpha = 0,1$ and for all other states they vanish in the asymptotic limit, $\gamma =1$. However, we find that the concurrence  vanishes for $\gamma \geq \gamma _{0}=1+\sqrt{4\alpha (1-\alpha )}-\sqrt{1+4\alpha (1-\alpha )}$. Thus,  unlike the earlier case, depolarizing  channel induces zero entanglement for $\gamma \geq  \gamma _{0}$, which is known as entanglement sudden death \cite{Werlang2009}. The critical value $\gamma _{0}$ vanishes for $\alpha =0,1$ (product states) with the maximum of $2-\sqrt{2}\approx 0.586$ for $\alpha =1/2$ (maximally entangled state). In Fig.\ref{fig2} we have shown entanglement sudden death for maximally entangled state. For comparison, we have also plotted MINs to show that quantum correlation exists even in the absence of entanglement. In other words, we conclude that MIN and F-MIN are more robust in quantifying the quantum correlation than the entanglement measure.    

\begin{figure*}[!ht]
\centering\includegraphics[width=0.49\linewidth]{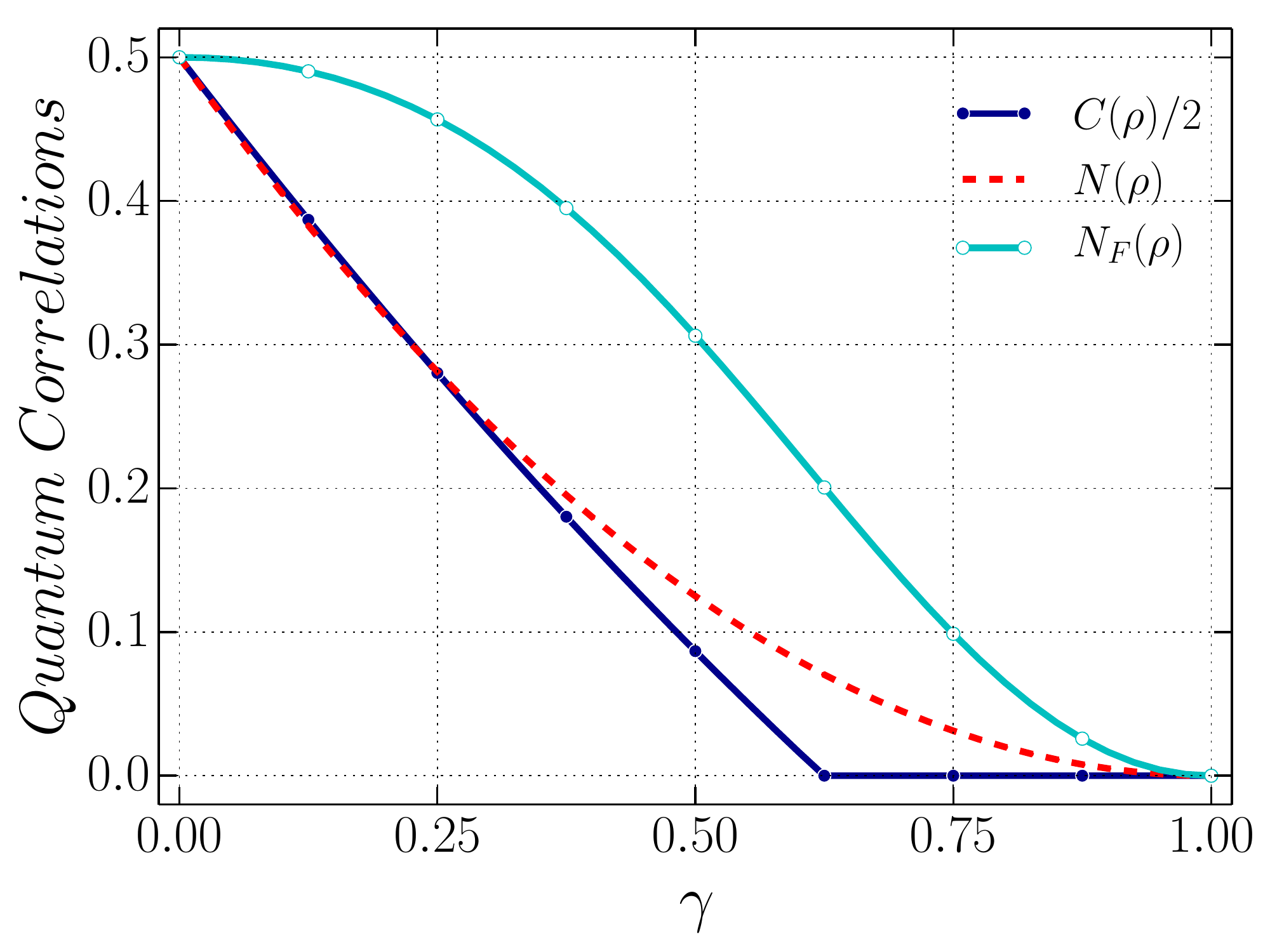}
\centering\includegraphics[width=0.49\linewidth]{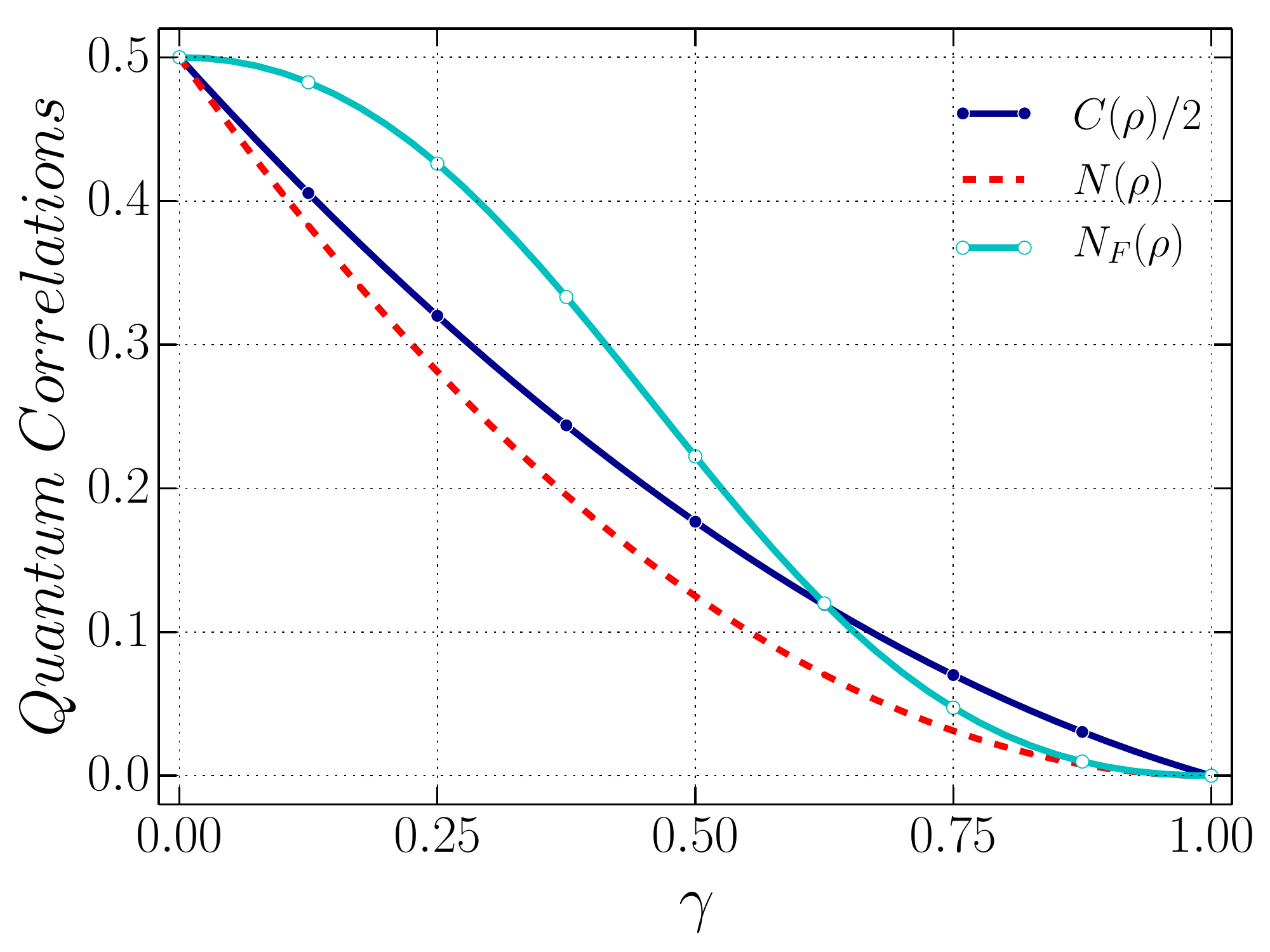}
\caption{(color online) Dynamics of concurrence, Hilbert- Schmidt norm and fidelity based measurement induced nonlocality under generalized amplitude damping channel for the initial state $| \psi \rangle =\frac{1}{\sqrt{2}}(| 00 \rangle + |11 \rangle )$ as as functions of $\gamma $ with $p=2/3$ (left) and $p=1$ (right).}
\label{fig3}
\end{figure*}

\textit{(iii) Generalized Amplitude Damping:}
Here we consider the generalized amplitude damping, which models the loss of energy from quantum system to environment at a finite temperature such as thermal bath. Such a process is described by the Kraus operators \cite{Nielsen2010}
\begin{eqnarray}
E_{0}&=&\sqrt{p}
\begin{pmatrix}
1 & 0\\
0 & \sqrt{1-\gamma }
\end{pmatrix},
~E_{1}=
\begin{pmatrix}
0 & \sqrt{\gamma }\\
0 & 0
\end{pmatrix}, \nonumber \\
E_{2}&=&\sqrt{1-p}
\begin{pmatrix}
\sqrt{1-\gamma} & 0\\
0 &  1
\end{pmatrix},
~E_{3}=\sqrt{1-p}
\begin{pmatrix}
0 & 0\\
 \sqrt{\gamma } & 0
\end{pmatrix}, \nonumber
\end{eqnarray} 
where $\gamma =1-\mathrm{e}^{-\gamma' t}$, $\gamma' $ is decay rate and $p$ defines the final probability distribution of stationary state. Non-zero matrix elements of evolved state under this channel are given by 
\begin{eqnarray}
\rho _{11}(t)&=&\rho _{11}(0) \{(1-\gamma) [2(1-p)-\gamma (1-2p)]\}+\gamma ^2 p^2,\nonumber \\
\rho _{22}(t)&=&\rho _{33}(t)=\gamma [\rho _{11}(0) (1-2p)(1-\gamma )+p(1-\gamma p)],\nonumber \\
\rho _{44}(t)&=&1-2\rho _{22}(t)-\rho _{11}(t),\nonumber \\
\rho _{14}(t)&=&\rho _{41}(t)=\rho _{14}(0)(1-\gamma).\nonumber
\end{eqnarray}
It is clear that the evolution of state in this channel depends on $\alpha $ and $p$. On substituting the evolved state elements in the expression of concurrence, we find that entanglement of the evolved state is zero for $\gamma \geq \gamma _0(\alpha ,p)$. To examine the dynamics, we consider the initial state $\alpha =1/2$. Setting $p=2/3$ we have $\gamma _{0}=0.6$. In other words, evolution of the maximally entangled state under this channel exhibits entanglement sudden death. On the other hand, $\gamma _0=1$ for $p=1$, implying that entanglement vanishes asymptotically. The dynamics of entanglement for the two cases are plotted in Fig.\ref{fig3} along with that of MINs. It is also clear from our results that, dynamics of both the MINs are qualitatively same. In particular, as time increases the MIN and F-MIN decrease showing that quantum correlation vanishes asymptotically. In this channel also the non-zero MINs in the region of zero concurrence show the existence of quantum correlation without entanglement.     
\section{Conclusions}
In this article, we have proposed measurement induced nonlocality (MIN) using fidelity induced metric as a measure of quantum correlation for bipartite state. It is shown that, in addition to capturing global nonlocal effect of a state due to von Neumann projective measurements, this quantity can be remedying local ancilla problem of MIN. We have presented a closed formula of fidelity based MIN for an arbitrary $2\times n$ dimensional mixed state, with an upper bound for $m\times n$ dimensional system. Further, we investigated the dynamics of proposed version of MIN under various noisy channel such as amplitude damping, depolarizing and generalized amplitude damping. Our results suggest that MIN and fidelity based MIN are more robust than entanglement against decoherence.


\bibliographystyle{99}

\begin{thebibliography}{26}

\bibitem{Ollivier2001}
H. Ollivier, W.H. Zurek, Phys. Rev. Lett. 88 (2001) 017901.

\bibitem{Girolami2011}
D. Girolami, G. Adesso, Phys. Rev. A 83 (2011) 052108.

\bibitem{Dakic2010}
Daki\'c, B. Vedral, V. Brukner, Phys. Rev. Lett. 105 (2010) 190502.

\bibitem{Luo2010pra}
S. Luo, S. Fu, Phys. Rev. A 82 (2010) 034302.

\bibitem{Luo2011}
S. Luo, S. Fu, Phys. Rev. Lett. 106 (2011) 120401.

\bibitem{Wiseman2007}
H.M. Wiseman, S.J. Jones, A. C. Doherty, Phys. Rev. Lett. 98 (2007) 140402.

\bibitem{Peters2005}
N.A. Peters, J.T. Barreiro, M.E. Goggin, T.-C. Wei, P.G. Kwiat, Phys. Rev. Lett. 94 (2005) 150502.

\bibitem{Mattle1996}
K. Mattle, H. Weinfurter, P.G. Kwiat, A. Zeilinger, Phys. Rev. Lett. 76 (1996) 4656.

\bibitem{Piani2012}
M. Piani, Phys. Rev. A 86 (2012) 034101.

\bibitem{Chang2013}
L. Chang, S. Luo, Phys. Rev. A 87 (2013) 062303.

\bibitem{Xi2012}
Z. Xi, X. Wang, Y. Li, Phys. Rev. A 85 (2012) 042325.

\bibitem{Hu2012}
M.L. Hu, H. Fu, Annals of Physics 327 (2012) 2343.

\bibitem{Li2016}
L. Li, Q.W. Wang, S.Q. Shen, M. Li, Europhys. Lett 114 (2016) 10007.

\bibitem{Hu2015}
M.L. Hu, H. Fan, New J. Phys. 17 (2015) 033004.

\bibitem{Rana2013}
S. Rana, P. Parashar, Quant. Inf. Process. 12 (2013) 2523.

\bibitem{Mirafzali2011}
S.Y. Mirafzali, I. Sargolzahi, A. Ahanj, K. Javidan, M. Sarbishaei, arXiv: 1110.3499v1 (2011).

\bibitem{Chen2015}
W.X. Chen, Y.X. Xie, X.Q. Xi, Int. J. Mod. Phys. B 29 (2015) 1550098.

\bibitem{Muthuganesan2017}
R. Muthuganesan, R. Sankaranarayanan Int. J. Mod. Phys 31 (2017) 1750166.

\bibitem{Sen2012}
A. Sen, D. Sarkar, A. Bhar , J. Phys. A: Math. Theor. 45 (2012) 405306.

\bibitem{Sen2013}
A. Sen, D. Sarkar, A. Bhar, Quant. Inf. Process 12 (2013) 3007.

\bibitem{Miszczak2009}
J.A Miszczak, Z. Pucha La, P. Horodecki, A. Uhlmann, K. Zyczkowski, Quant. Inf. Comput 9 (2009) 0103.  

\bibitem{Jozsa1994}
R. Jozsa, Journal of Modern Optics 41 (1994) 2315.

\bibitem{Wang2008}
X. Wang, C.S. Yu, X.X. Yi, Physics Letters A 373 (2008) 58.

\bibitem{SLuo2012}
S. Luo, S. Fu, Theor. Math. Phys. 171 (2012) 870.

\bibitem{Luo2010}
S. Luo, S. Fu, Europhys. Lett 92 (2010) 20004.

\bibitem{Horodecki1999}
M. Horodecki, P. Horodecki, Phys. Rev. A 59 (1999) 4206.

\bibitem{Werner1989}
R.F. Werner, Phys. Rev. A 40 (1989) 4277.

\bibitem{Nielsen2010}
M. Nielsen, I. Chuang, Quantum Computation and Quantum Information, Cambridge (2010).

\bibitem{Hill1997}
S. Hill, W.K. Wootters, Phys. Rev. Lett. 78 (1997) 5022.

\bibitem{Werlang2009}
T. Werlang, S. Souza,  F.F. Fanchini, C.J. Villas Boas, Phys Rev. A 80 (2009) 024103.

\end{thebibliography}

\end{document}